\documentclass[%
aps,prapplied,twocolumn,
superscriptaddress,
amsmath,amssymb,
floatfix,]{revtex4-2}
\usepackage{graphicx}
\usepackage{dcolumn}
\usepackage{bm}
\usepackage{multirow}
\usepackage{braket}
\usepackage{amsfonts}
\usepackage{amssymb}
\usepackage{amsthm}
\usepackage{array}
\usepackage{amsmath}
\usepackage{verbatim}
\usepackage{hyperref}
\usepackage{color}
\usepackage{bbold}
\usepackage{epstopdf}
\usepackage{mathtools}
\usepackage{MnSymbol}
\usepackage{tikz}
\usepackage[export]{adjustbox}
\usepackage{subfigure}
\usepackage{threeparttable}
\usepackage{hyperref}
\usepackage{xcolor}

\newcommand{\affa}{Beijing Key Laboratory of Fault-Tolerant Quantum Computing, Beijing Academy of Quantum Information Sciences, Beijing 100193, China}

\newcommand{\affb}{Beijing National Laboratory for Condensed Matter Physics, Institute of Physics, Chinese Academy of Sciences, Beijing 100190, China}
\newcommand{\affc}{University of Chinese Academy of Sciences, Beijing 100049, China}

\newcommand{\afff}{Hefei National Laboratory, Hefei 230088, China}

\newcommand{\affg}{School of Physics and State Key Laboratory of Nuclear Physics and Technology, Peking University, Beijing 100871, China}

\newcommand{\affh}{Institute of Theoretical Physics, Chinese Academy of Sciences, Beijing 100190, China}

\newcommand{\affi}{School of Physical Sciences, University of Chinese Academy of Sciences, Beijing 100049, China}

\newcommand{\abs}[1]{\left|#1\right|}

\def\bra#1{\langle #1|}
\def\ket#1{\left|#1 \right>}

\begin{document}
	
        \title{Characterizing charge-parity detection based on an offset-charge-tunable transmon qubit via randomized benchmarking}
	\author{Yao-Yao Jiang}
    \thanks{These authors contributed equally to the work.}
	\affiliation{\affa}
	\affiliation{\affb}
	\affiliation{\affc}
    \author{Tang Su}
    \thanks{These authors contributed equally to the work.}
	\affiliation{\affa}
    \author{Yuxiang Liu}
    \affiliation{\affg}
    \author{Yi-Ming Guo}
	\affiliation{\affa}
	\affiliation{\affb}
	\affiliation{\affc}
    \author{Yidong Song}
    \affiliation{\affh}
    \affiliation{\affi}
    \author{Yu-Long Li}
	\affiliation{\affa}
    \author{Yanjie Zeng}
    \affiliation{\affh}
    \affiliation{\affi}
	\author{Guang-Ming Xue}
	\affiliation{\affa}
	\affiliation{\afff}
	\author{Wei-Jie Sun}
	\affiliation{\affa}
    
    \author{Mei-Ling Li}
	\affiliation{\affa}
    \author{Yi-Rong Jin}
	\affiliation{\affa}
    \affiliation{\afff}
    \author{Junhua Wang}
    \email{wangjh@baqis.ac.cn}
	\affiliation{\affa}
    \author{Xuegang Li}
    \email{lixg@baqis.ac.cn}
    \affiliation{\affa}
	\author{Hai-Feng Yu}
	\email{hfyu@baqis.ac.cn}
	\affiliation{\affa}
	\affiliation{\afff}

\begin{abstract}

Superconducting qubits are compelling platforms for charge-parity detection and, due to their theoretical sensitivity on the meV energy scale, hold promise for rare event searches.
In this work, we realize high-fidelity mapping of charge-parity states onto qubit states using an offset-charge-tunable transmon qubit and efficiently characterize the fidelity of the charge-parity detection via randomized benchmarking. Specifically, a gate control line is applied to control offset charge, allowing us to achieve the single-qubit gate fidelity up to 99.96\%. We combine a net-zero-based pulse on the gate line with a spin-echo-based sequence to realize charge-parity mapping, achieving a fidelity of 99.37\%. Then, we demonstrate continuous monitoring of the charge-parity state with over 93.4\% fidelity at a 4-$\mu$s sampling interval. Finally, an error analysis of charge-parity detection is performed, and it is found that qubit readout is currently the largest source of error. We believe this work lays the foundation for future exploration of ultra-low energy particles.

\end{abstract}
    \maketitle
    \bigskip
    \section{INTRODUCTION}
    
    Superconducting quantum computing has advanced rapidly, characterized by a significant increase in the number of physical qubits~\cite{AbuGhanem2025,jin2025topological,gao2025establishing}, improvements in qubit gate fidelity~\cite{mckay2023, Fabian2025,li2023error}, and the successful demonstration of quantum error correction~\cite{Acharya2024,Krinner2022,Zhao2022}. However, these superconducting qubits are susceptible to quasiparticle (QP) bursts, for example from cosmic rays~\cite{McEwen_2021, harrington2025, Li2025,Wilen_2021}. Theoretically, superconducting qubits are sensitive to even energy deposition at the meV scale, an amount sufficient to break Cooper pairs in the superconductor and generate QPs. These QPs can tunnel across the Josephson junctions of the qubit and thus induce the charge-parity switches and qubit state transitions~\cite{catelani2011,catelani2014parity,Rist2013,Serniak_2018}. This presents both challenges and opportunities. On one hand, it requires complex shielding~\cite{Bertoldo2025,loer2024abatement,bratrud2024first} and error mitigation protocols~\cite{Iaia_2022, Xu_2022,mcewen2024,wu2025mitigating} to ensure quantum computational robustness; on the other hand, it also makes superconducting qubits a promising detector for single far-infrared photons, dark matter particles, and neutrinos~\cite{chou2023,ramanathan2024,fink2024,Linehan2025}. 

    Theoretical schemes for low-energy particle detection based on charge-parity switches of superconducting qubits have been proposed, known as the quantum parity detector (QPD)~\cite{ramanathan2024, fink2024}. 
    Experimentally, quantum capacitance detectors based on charge-sensitive superconducting qubits~\cite{Nakamura1999}(a type of QPD) have been shown to be able to detect single far-infrared photons at 1.5 THz~\cite{echternach2018}. The key point of these schemes is to detect charge-parity states of qubits.

    A straightforward strategy is to read out the parity-dependent qubit frequency shift by coupling the qubit to a microwave waveguide~\cite{fink2024, amin2024}. While this approach is convenient to implement, it suffers from slightly lower readout photon-collection efficiency. To address this, one can introduce a resonator as an intermediary~\cite{Serniak2019, ramanathan2024}, collecting more coherent signal photons. Ref.~\cite{Serniak2019Parity} has achieved a detection fidelity of 99\% using direct dispersive readout based on the parity-dependent resonator frequency shift. However, direct dispersive charge-parity measurement may also increase the probability of readout-induced transitions in qubits~\cite{Sank2016,Khezri2023,Nesterov2024,fechant2025offset,connolly2025full}.  
    Reducing the dispersive coupling strength between the qubit and the resonator can help mitigate such errors, but at the cost of greatly decreasing the parity-dependent resonator frequency shift. In this regime, mapping the charge-parity states to the qubit states, followed by the qubit dispersive readout, is a more convincing approach~\cite{Rist2013,Serniak_2018}.
    
    Refs.~\cite{Rist2013,Serniak_2018} utilize Ramsey-based pulse sequences to detect the parity-dependent qubit frequency shift and demonstrated a charge-parity detection fidelity of 91-92\%. However, when scaled to larger two-dimensional QPDs, this method may be more vulnerable to low-frequency noise. Furthermore, to our knowledge, an randomized-benchmarking-like (RB-like) protocol to quantify the average error rate of charge-parity detection is still lacking.

    \begin{figure*}[t]
    \includegraphics[width=0.9\textwidth]{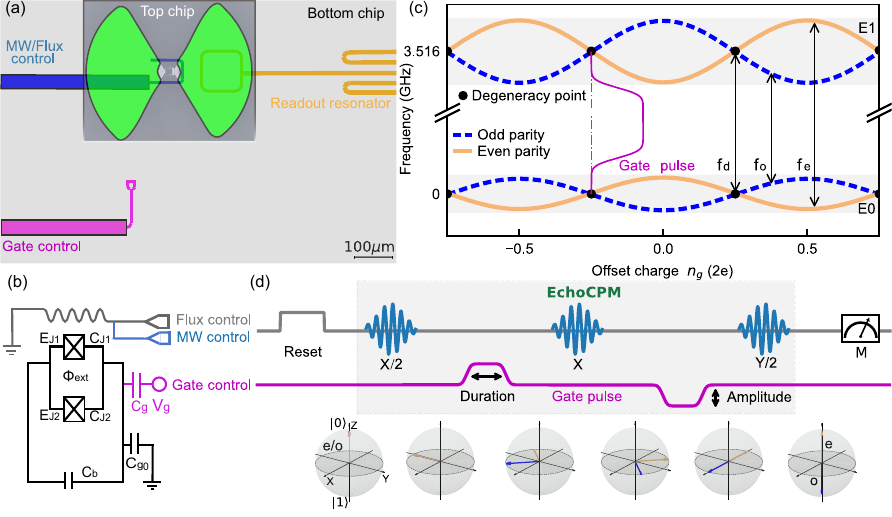}
    \caption{{(a)} False-colored optical image of a single-qubit unit in the device. The bottom chip comprises a microwave (MW) or flux control line (blue), a gate control line (magenta), and a $\lambda/4$ readout resonator (orange). The top chip contains only a transmon qubit (green). {(b)} The circuit diagram of the device shown in {(a)}. {(c)} The two lowest energy levels ($E_0$ and $E_1$) of the qubit as a function of the offset charge $n_g(2e)$ for even (solid line) or odd (dashed line) charge-parity states. The parity-dependent qubit frequencies ($f_e$ and $f_o$) are degenerate at $n_g = 0.5n+0.25, n=0,\pm1,\pm2,...$ (degeneracy points). The $n_g$ can be tuned away and back to the degeneracy point with a gate pulse (magenta) to accumulate parity-dependent phases of the qubit. {(d)} Spin-echo-based sequence with an inserted net-zero-based gate pulse for charge-parity mapping, named EchoCPM, where several Bloch sphere diagrams show corresponding qubit states. Charge-parity detection can be realized by first resetting the qubit, performing the EchoCPM, and then following the qubit measurement.}
    \label{fig:Figure_1}
\end{figure*}

In this work, we implement a high-fidelity charge-parity mapping based on a spin-echo-like sequence~\cite{Hahn1950}, named EchoCPM, and characterize its performance via Clifford-based randomized benchmarking (RB)~\cite{knill2008randomized,Magesan2011refRB}. First, we add a gate line to control the offset charge of a transmon qubit, enabling fast offset-charge calibration. Then, by integrating a net-zero-based pulse~\cite{Negrneac2021} on the gate line (gate pulse) with spin-echo-based sequences (see Fig.~\ref{fig:Figure_1}(d)), we can accumulate a $\pi$ phase difference between two charge-parity states, creating a robust charge-parity mapping framework. By constructing this mapping into a Clifford Z gate, we can measure its fidelity using RB. Combined with qubit state preparation and measurement, we demonstrate continuous charge-parity detection and provide a comprehensive error analysis. 
    
	
    \bigskip

    
    \section{Device and charge-parity detection}

     In our experiment, we employ an offset-charge-tunable transmon qubit as a QPD. As depicted in Fig.~\ref{fig:Figure_1}(a), the device uses a flip-chip layout comprising a top qubit chip and a bottom carrier chip. The qubit chip hosts solely a transmon qubit, while the carrier chip integrates a readout resonator, a microwave (MW) or flux control line, and a gate control line. The MW/flux control line, which is inductively coupled to the qubit, is used to drive the qubit and tune its frequency. The gate control line, which is capacitively coupled to the qubit, is used to tune the offset charge. This design can be easily scaled up to large two-dimensional QPD arrays.

     Fig.~\ref{fig:Figure_1}(b) shows the quantum circuit of the QPD. The system Hamiltonian can be described as follows~\cite{Koch_2007}, 
    \begin{equation}
    \begin{aligned}
    H/\hbar = &\omega_{q} a^\dag a - \dfrac{\eta}{2}a^\dag a^\dag a a
    \\ &- \dfrac{1}{2}\sum_{k}\epsilon_{k}\cos(2\pi (n_g-\dfrac{P-1}{4}))\ket{k}\bra{k},
    \end{aligned}
    \end{equation}
     where $\omega_{q}$ and $\eta$ are the frequency and the anharmonicity of the qubit, $a^{\dag}$ and $a$ are the raising and lowering operators, $n_g(2e)$ is the offset charge, $P=1$ ($P=-1$) is the odd (even) charge parity of the qubit, $\epsilon_{k}$ is the charge dispersion of the $k$-th energy level. In Fig.~\ref{fig:Figure_1}(c), we provide a sketch of the two lowest energy levels for the qubit. The cosinoidal dispersions of different charge-parity states cross at $n_g = 0.5n+0.25, n=0,\pm1,\pm2,...$, defined as the degeneracy points, where we can apply a MW drive pulse with a frequency of $f_d$ to enable high-fidelity single-qubit gate operations. Away from the degeneracy point, the system exhibits two separate qubit frequencies, $f_e$ (even) and $f_o$ (odd). In this off-degeneracy regime, we can realize parity-dependent phase accumulations.
     
     Fig.~\ref{fig:Figure_1}(d) shows the measurement sequence of charge-parity detection. First, we reset the qubit and bias it at the degeneracy point. Then we apply a spin-echo-based sequence (X/2-X-Y/2)~\cite{Hahn1950} on the MW/flux control line instead of the conventional Ramsey-based sequence to detect the qubit frequency shift~\cite{Rist2013}. This can maintain robustness to low-frequency noise. To distinguish two charge-parity states, we insert two gate pulses of equal magnitude and opposite signs. One pulse is placed between the X/2 and X gates, and the other between the X and Y/2 gates. These pulses displace the qubit from the degeneracy point, generating opposite frequency shifts for the two parity states. This net-zero-based approach improves distortion invariance~\cite{Negrneac2021}. The insertion of the X gate adds up the phases accumulated by the first and the second gate pulse rather than canceling. Once the gate pulse is precisely calibrated, a total phase difference $\delta = \pi$  between the two charge-parity states can be achieved. Followed by a Y/2 gate, we can realize a high-fidelity mapping of charge-parity states to qubit states. Finally, the charge-parity detection is completed by measuring the qubit states.

\begin{flushleft}    
\begin{figure}[t]
    \includegraphics[width=\linewidth]{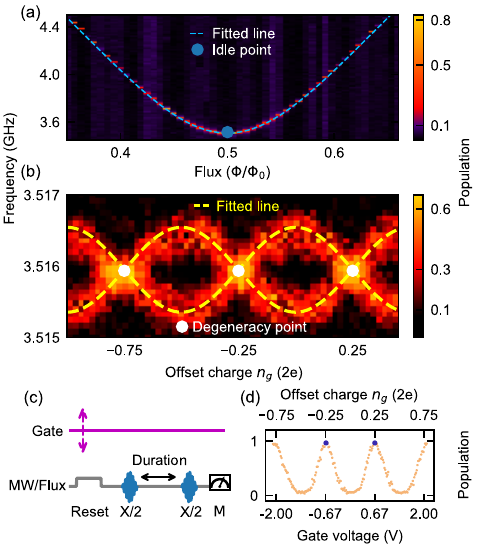}
    \caption{{(a)} Qubit spectrum as a function of flux $\Phi/\Phi_0$, where $\Phi_0$ is the single flux quantum. The flux point with the lowest qubit frequency is chosen as the idle point, featuring the best coherence. The blue dashed line is a fit based on the transmon Hamiltonian. {(b)} Qubit spectrum as a function of the offset charge $n_g(2e)$. Two traces for two charge-parity states are fitted to give the charge dispersion energy difference. Their cross points (white dots) are identified as the degeneracy points. {(c)} The Ramsey-based sequence with a duration of 800 ns to rapidly and precisely identify the degeneracy point. {(d)} The qubit population as a function of the gate voltage, with the qubit drive frequency set at the degeneracy point. The maximum points (blue dots) are identified as the degeneracy points.}
    \label{fig:Figure_2}
    
\end{figure}
\end{flushleft}
    \bigskip
    \section{Identification of the degeneracy point}

    Our experiment utilizes an asymmetrical Josephson junction qubit, exhibiting a spectrum with higher and lower sweet points at flux $\Phi/\Phi_0 = 0$ and $0.5$, respectively, where $\Phi_0$ is the flux quantum~\cite{Koch_2007}. As shown in Fig.~\ref{fig:Figure_2}(a), the lower sweet point (blue dot) serves as the high-coherence idle point, near which the qubit operates at 3.516 GHz (see Tab.~\ref{tab:simulation_parameters} in Appendix C for detailed parameters). We can fast reset the qubit by tuning the qubit frequency to match the 6.3-GHz frequency of the readout resonator. With the qubit fixed at 3.516 GHz, we measure the qubit spectrum as a function of the offset charge, as shown in Fig.~\ref{fig:Figure_2}(b). The resulting qubit spectrum exhibits two distinct branches, which correspond to the even and odd charge-parity states, respectively. The degeneracy points are identified at the crossing points (white dots). We fit the data to $\epsilon_{10} \cos(2\pi(n_g - (P-1)/4))$, where $\epsilon_{10}=\epsilon_1-\epsilon_0$, and achieve a charge dispersion energy difference $\epsilon_{10}$ of -1.192 MHz.
    
    To rapidly and accurately identify the degeneracy point, we utilize a Ramsey-based sequence to monitor the offset charge~\cite{Wilen_2021}, as illustrated in Fig.~\ref{fig:Figure_2}(c). Given that $\epsilon_{10} = -1.192$ MHz and aiming to maximize the sensitivity, we set the duration between the two X/2 pulses to 800 ns. With this setup, the measured population is maximal at the degeneracy point and minimal when $n_g$ is an integer or half-integer (charge sweet point), as shown in Fig.~\ref{fig:Figure_2}(d). Thus, we can select one of the maxima as the degeneracy point to maintain optimal qubit performance. This procedure is calibrated every $\sim5$ min due to offset charge fluctuation, and each recalibration can be completed within few seconds.
	
    \bigskip
    \section{Fidelity of charge-parity mapping}

\begin{flushleft}          
\begin{figure}[t]
    \centering
    \includegraphics[width=1\linewidth]{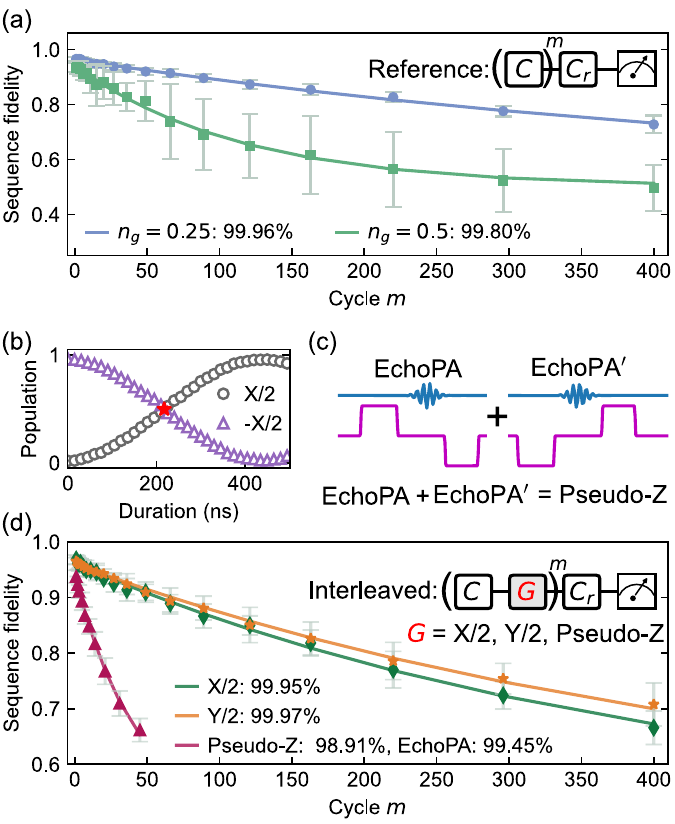}
    \caption{{(a)} Single-qubit randomized benchmarking (RB) at $n_g = 0.25$ (circles) and  $n_g = 0.5$ (squares). After power-law fitting, we achieve average single-qubit-gate fidelities of 99.96$\%$ and 99.8$\%$ for $n_g = 0.25$ and $n_g = 0.5$, respectively. Error bars denote the standard deviation. The inset depicts the reference RB sequence. {(b)} Duration calibration of the net-zero-based gate pulse. The pulse sequence is the same as Fig.~\ref{fig:Figure_1}(d) except that the Y/2 gate is replaced by the X/2 (circles) and -X/2 (triangles) gates. The cross point (star) with a duration of 217 ns is selected as the working point. {(c)} A pulse sequence consisting of two opposite-sign, echo-based phase accumulation (echoPA and echoPA$^\prime$) is used as the Z gate, and is called a pseudo-Z gate. {(d)}  Interleaved RB for three gates—X/2 (diamonds), Y/2 (stars), and pseudo-Z (triangles). Power-law fittings using the RB in (a) as reference yield gate fidelities of 99.95$\%$ (X/2), 99.97$\%$ (Y/2), and 98.91$\%$ (pseudo-Z). Error bars denote the standard deviation. The inset shows the interleaved RB sequence. The single echoPA fidelity is $\sqrt{98.91\%} = 99.45\%$ and the fidelity of the charge-parity mapping is 99.45$\%\times$99.95$\%\times$99.97$\%$=99.37$\%$.}
    \label{fig:Figure_3}
\end{figure}
\end{flushleft}

    
    In this work, we employ a 20-ns MW drive pulse, equipped with a cosine envelope and a 5-ns buffer, to realize arbitrary single-qubit gates using the phase-shifted microwave method~\cite{Chen2023}. We utilize the derivative removal by adiabatic gate to suppress the leakage errors~\cite{Motzoi2013}.
    We then assess the average fidelity of the single-qubit gate using the Clifford-based RB sequence, as shown in the inset of Fig.~\ref{fig:Figure_3}(a). After the qubit initialization, $m$ random single-qubit Clifford group elements $\boxed{C}$
    and their inverses $\boxed{C_{r}}$ are applied sequentially, followed by the qubit measurement. The sequence fidelity as a function of cycle $m$ follows $Ap_{\rm{ref}}^m+B$, where $A$ and $B$ capture the state preparation and measurement errors, $p_{\rm{ref}}$ is the sequence decay parameter. Using this framework, we achieve an averaged single-qubit-gate fidelity of 99.96\% when $n_g=0.25$ and 99.80\% when $n_g=0.5$. The single-qubit gate error of the former is nearly an order of magnitude lower than that of the latter, and also shows an obviously smaller variation. 

    Fig.~\ref{fig:Figure_3}(b) demonstrates a parity-independent calibration process designed to achieve the phase difference $\delta=\pi$ for charge-parity mapping. The pulse sequence employed herein is the same as the EchoCPM, with the minor modification of replacing the Y/2 gate with the X/2 and -X/2 gates. To enable fast phase accumulation, we select a smoothed square gate pulse. With a fixed gate pulse amplitude of $\sim$0.67 V, which can tune the offset charge from the degeneracy point to near the charge sweet point, we measure the population as a function of the gate pulse’s duration. The two population traces intersect at $\delta=\pi$ and show the pulse duration of 217 ns (star), near the theoretical value $1/4\abs{\epsilon_{10}}\sim210$ ns. We set this value to achieve the charge-parity mapping. Because the charge-parity-mapping is not a quantum gate, we decompose it into three parts: X/2, Y/2, and a central block with a spin-echo-based phase accumulation (echoPA) as shown in Fig.~\ref{fig:Figure_3}(c). An echoPA is represented as $R_Z(\pm\delta/4)R_X(\pi)R_Z(\mp\delta/4)$, with the sign `+' or `-' depending on the even or odd parity state, where $R_\alpha(\theta)$ denotes a rotation about the axis $\alpha$ by an angle of $\theta$. We further construct a composite sequence by concatenating an echoPA with another pulse called echoPA$^\prime$, where the latter’s gate pulse has the opposite sign to echoPA. This sequence can be expressed as:
    \begin{equation}
    \begin{aligned}
    &R_Z(\pm\frac{\delta}{4})R_X(\pi)R_Z(\mp\frac{\delta}{4})R_Z(\mp\frac{\delta}{4})R_X(\pi)R_Z(\pm\frac{\delta}{4})
    \\&=R_Z(\pm\delta)
    \end{aligned}
    \end{equation}
    When $\delta=\pi$, this yields an equivalent Z gate, which is independent of charge parity, and is named the pseudo-Z.

    As shown in Fig.~\ref{fig:Figure_3}(d), we individually benchmark the X/2, Y/2, and pseudo-Z gates using interleaved RB~\cite{magesan2012efficient}. The inset shows the sequence where each target gate is interleaved in the RB sequence. Fitting the sequence fidelity as a function of cycle $m$ with a power law yields the decay parameter $p_{\rm{int}}$. For a single qubit, following $1-(1-p_{\rm{int}}/p_{\rm{ref}})/2$, we obtain interleaved gate fidelities of 99.95$\%$, 99.97$\%$, and 98.91$\%$ for the X/2, Y/2 and pseudo-Z, respectively. For each echoPA, the fidelity can be calculated as $99.45\%$. Finally, the fidelity of the charge-parity mapping is $99.45\%\times99.95\%\times99.97\%=99.37\%$. We also simulate the process of charge-parity mapping with Qsim~\cite{quantum_ai_team_and_collaborators_2020_4023103} and achieve a fidelity of 99.36\%, indicating the fidelity of the charge-parity mapping is only limited by decoherence. 

    \bigskip
    \section{Fidelity of charge-parity detection}
    
\begin{figure}[t]
    \noindent\includegraphics[width=\linewidth]{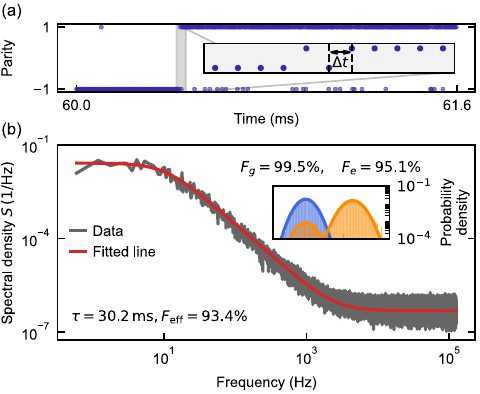}
    \caption{{(a)} A 1.6-ms time slice of the charge-parity detection trace shows a charge-parity switching event. The event is zoomed in the inset, which shows that the time interval between adjacent data points is $\Delta t = 4~\mu$s. 
    {(b)} Power spectral density of the detection trace. The red curve represents the Lorentzian fit. Inset: histograms of the qubit readout distribution for ground and excited states. }
    \label{fig:Figure_4}
    \end{figure}

    We demonstrate real-time charge-parity detection by continuously executing the sequence in  Fig.~\ref{fig:Figure_1}(d) over a time duration of 30 s. Fig.~\ref{fig:Figure_4}(a) shows a 1.6-ms segment of the charge-parity measurement time trace, recorded with a sampling period of $\Delta t=4$ $\mu$s. The observed charge-parity switch is primarily due to the tunneling of residual QPs  \cite{Rist2013,Li2025,Serniak_2018}. The result can be well described by a random-telegraph signal (RTS). Fourier transform of the autocorrelation of the RTS, taking into account qubit readout fidelities ($F_g$ and $F_e$), yields a Lorentzian power spectral density (PSD) as,
    \begin{equation}
    \label{eq:Sf-Feff-final}
    S(f_n)=
    \frac{F_{\mathrm{eff}}^{\,2}\,4\Gamma}{(2\Gamma)^{2}+(2\pi f_n)^{2}}
    \;+\;\Bigl[1-F_{\mathrm{eff}}^{\,2}-(F_e-F_g)^{2}\Bigr]\Delta t, 
    \end{equation}
    where 
    $1/\Gamma=\tau$ is the average QP-tunneling time, $F_{\rm{eff}} = (F_g+F_e-1)F_m$ is the effective fidelity of charge-parity detection, $F_m$ is the fidelity of the charge-parity mapping (see Appendix D for more details). As shown in Fig.~\ref{fig:Figure_4}(b), the PSD of our result is fitted (red line) and gives $F_{\rm{eff}} \,(\rm{fitted}) = 93.4\%$, $\tau=30.2$ ms. In our experiment, $F_g = 99.5\%$ and $F_e = 95.1\%$ and $F_m = 99.37\%$. This gives $F_{\rm{eff}} \,(\rm{calculated}) = 94.0\%$, which is approaching $93.4\%$, indicating the fidelity of the charge-parity detection in our experiment is primarily limited by the qubit readout. The measure QP-tunneling time $\tau$ is shorter than that measured in the Refs.~\cite{Pan2022,Connolly2024,Yelton2025}, we attribute this difference to the leakage of infrared photons and stress release events in our device.
    
	\bigskip
    \section{DISCUSSION}
        
        In conclusion, we employ gate voltage control on an offset-charge-tunable transmon qubit, achieving single-qubit gates with a fidelity of 99.96\% and charge-parity mapping with a fidelity of 99.37\%. Additionally, we demonstrate continuous charge-parity detection over a time interval of 4 $\mu$s, which yields a fidelity of 93.4\%. Error analysis indicates the qubit readout as the key factor limiting this performance. Given that state-of-the-art superconducting-qubit readout fidelity has been reaching 99.9\%~\cite{Fabian2025,wang2025USTC}, we anticipate that enhancing readout performance will greatly improve charge-parity detection fidelity. Furthermore, our scheme exhibits noise robustness and can be easily extended to larger two-dimensional QPD arrays. Therefore, we believe our work paves the way for low-energy-threshold detectors based on superconducting qubits.

\bigskip
\noindent\textbf{DATA AVAILABILITY}

Source data files are available at \hyperlink{https://doi.org/10.6084/m9.figshare.30636059}{https://doi.org/10.6084/m9.figshare.30636059}, and other data are available from the corresponding author upon request.

\bigskip
\noindent\textbf{ACKNOWLEDGMENTS}

We acknowledge supports from the National Natural Science Foundation of China (Grant Nos. 92365206 and 12441504), and Innovation Program for Quantum Science and Technology (No.2021ZD0301802).

\bigskip



\appendix

\section{Experiment setup}

\begin{figure*}[t]
\centering
\includegraphics[width=0.6 \textwidth]{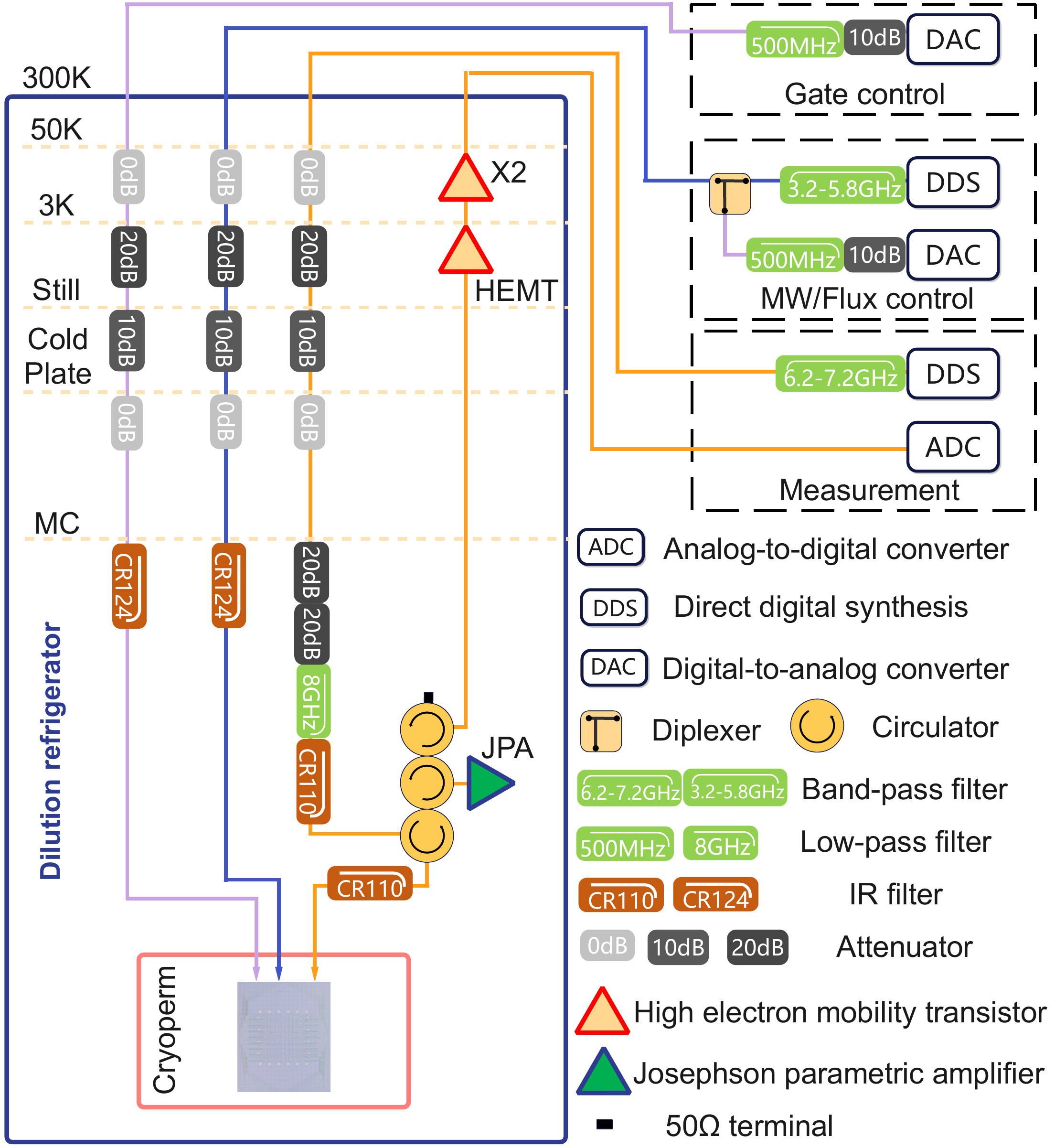}
\caption{{\bf Schematic of the measurement setup.}}
\label{fig:Measurement setup}

\end{figure*}

\subsection{Device design}
In this work, we employ an offset-charge-tunable transmon qubit as the charge-parity detector. To achieve full control over the qubit, we utilize two dedicated control lines: a microwave (MW)/flux control line and a gate control line. The MW/flux line is inductively coupled to the qubit, allowing us to apply MW
drives to prepare qubit states or tune the qubit's frequency (using static bias or fast flux pulses).  The gate control line is capacitively coupled to the qubit, serving dual purposes: stabilizing the offset charge and enabling biasing the qubit away from the charge-degeneracy point to realize a fast gate pulse. For qubit state measurement, we couple the qubit to a $\lambda/4$ resonator, which is read out in reflection mode. The resonator is set to a frequency of 6.3 GHz—a value intentionally chosen to lie between the qubit’s tunable frequency range. For qubit reset, we bias the qubit into resonance with the resonator. Note that the resonator has a linewidth of $\kappa \approx 0.7$ MHz, which serves as a dissipation channel. This channel facilitates rapid energy exchange between the qubit and the resonator, ultimately enabling the qubit to be rapidly reset. Additionally, all control lines and transmission lines in the system employ a tunnel-type air-bridge crossover. 

\subsection{Fabrication }


The device adopts a two-chip architecture. The top chip contains the qubit devices, including the capacitor pads of the qubit and Josephson junctions, while the bottom chip carries control lines, a readout resonator, and transmission lines. The fabrication of the top chip proceeds in two main steps. First, a tantalum film is deposited on a sapphire substrate, patterned by photolithography, and etched using inductively coupled plasma (ICP) to define the capacitor pads of the qubit. Second, the Josephson junctions are patterned by electron-beam lithography with bilayer resists and fabricated using the standard Dolan-bridge shadow evaporation
technology. For the fabrication of the bottom chip, we deposit a tantalum film on a silicon substrate. All kinds of wiring lines are defined via photolithography and then etched using ICP to create the final conductive patterns. Then, two additional photolithography steps are followed by aluminum deposition and wet etching to form tunnel-type air-bridges. To further remove the photoresist, an ozone treatment is performed after the lift-off. Then, 15 $\mu$m-thick photoresist layer is spin-coated onto the bottom chip and patterned by photolithography to define the indium bumps. Next, a 10 $\mu$m-thick indium film is deposited by thermal evaporation, and then the sample is immersed in N-methyl-2-pyrrolidone to lift-off the photoresist. Finally, the two chips are aligned and bonded using flip-chip technology. To enhance the mechanical robustness of the inter-chip connection, low-temperature glue is applied along the corners of the chips. 



\subsection{Measurement setup }




Our measurement setup is shown in Fig.~\ref{fig:Measurement setup}. 
We use the single-control-line design to realize the simultaneous XY drive and Z flux bias for the qubit. The XY drive signal is generated by a direct digital synthesizer (DDS) with a sampling rate of 6\,GSa/s and filtered by a 3.2–5.8\,GHz band-pass filter. The Z flux bias signal is generated by a digital-to-analog converter (DAC) with a sampling rate of 2.5\,GSa/s and processed by a 10\,dB attenuator and a 500 MHz low-pass filter. The two signals are combined at room temperature by a diplexer and delivered to the qubit through the MW/flux single-control line. The gate tuning signal is generated by a DAC with a sampling rate of 2.5\,GSa/s, processed by the same attenuator and filter as flux bias signals, and delivered to the qubit through the gate control line.

For qubit readout, the drive signal is generated by a DDS with a sampling rate of 6\,GSa/s, and filtered by a 6.2–7.2\,GHz band-pass filter. This signal is then gradually attenuated before being applied to the readout resonator. The reflected signal, which encodes the qubit’s state information, subsequently passes through a circulator, Josephson parametric amplifier (JPA) and a isolator. This signal is then amplified by a high electron mobility transistor (HEMT) at the 3\,K stage, and further amplified by two HEMTs at the 50\,K stage. We capture this signal by an analog-to-digital converter (ADC) with a sampling rate of 4\,GSa/s.

To suppress infrared photons from cables and environment, infrared filters (Eccosorb\textsuperscript{\textregistered} CR110 and CR124) are integrated into each control line. This ensures that extraneous infrared radiation is attenuated before it can interfere with the qubit. Additionally, the sample box is reinforced to enhance its sealing performance, further minimizing the ingress of environmental infrared photons. These combined strategies effectively reduce the quasiparticle background in the system. In our experiment, we achieve an average background quasiparticle tunneling time of approximately 32 ms. In the future, we believe an even lower quasiparticle background can be achieved via improved device design and enhanced sealing strategies.

\section{Interleaved-RB-Based Optimization of the gate Pulse duration}

\begin{figure}
    \centering  \includegraphics[width=1\linewidth]{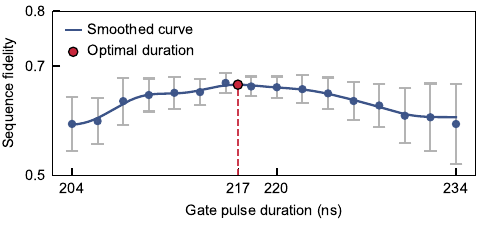 }
    \caption{Optimization of the gate-pulse  duration via interleaved-RB. Dark blue dots represent the mean of sequence fidelity measured with cycle length of $m=50$. Gray error bars denote the standard deviation. The dark blue line represents the smoothed curve, where the peak (red dot) indicates the optimal duration of 217 ns. 
    }
    \label{fig:rb_opt}
\end{figure}

We employ interleaved randomized benchmarking (interleaved RB) to optimize the gate-pulse parameter. The gate-pulse amplitude is fixed at $\sim0.67\,\mathrm{V}$, and the pulse duration is scanned within a narrow window around a preselected reference of $220\,\mathrm{ns}$. For each pulse duration, we perform an interleaved RB experiment by inserting a pseudo-Z gate into random Clifford sequences, with the interleaved-RB cycle length fixed at $ m=50$. The sequence fidelity serves as the optimization metric. Maximizing this metric yields the optimal duration $217\,\mathrm{ns}$, as shown in Fig.~\ref{fig:rb_opt}. To avoid collecting erroneous signals due to the slow drift of the offset charge, we execute a degeneracy point calibration sequence immediately before and after each interleaved-RB experiment. We only retain datasets where the gate voltage values measured at the degeneracy point differ by no more than $\pm 0.1\,\mathrm{V}$. 


\section{Simulation of Charge-Parity Mapping}



We numerically simulate the charge-parity mapping on an offset charge-tunable transmon qubit using the Qsim software. The system Hamiltonian can be written as,
\begin{equation}
\begin{aligned}
\label{eq:Ht-full}
\frac{H_0}{\hbar}
&=\ \omega_q\, a^\dagger a
\;+\; \frac{\eta}{2}\, a^{\dagger 2} a^2 \\
\frac{H(t)}{\hbar}
&=
\underbrace{\frac{H_0}{\hbar}}_{\text{static}}
\;+\;
\underbrace{\Bigg[
-\frac{1}{2}\sum_{k\ge 0}\frac{\epsilon_k}{\hbar}\,
\cos\!\Big(2\pi\big[n_g(t)+\tfrac{P-1}{4}\big]\Big)\,|k\rangle\langle k|
\Bigg]}_{\text{gate drive (via }n_g(t)\text{)}}\\
\;&+\;
\underbrace{\!\left[\Omega_{\mathrm{WM}}(t)\,a^\dagger+\Omega_{\mathrm{WM}}^*(t)\,a\right]}_{\text{WM drive}},
\end{aligned}
\end{equation}
where $a^\dagger$ and $a$ are bosonic creation and annihilation operators, $\omega_q$ is the qubit frequency, $\eta$ is the anharmonicity of the qubit, $\epsilon_k$ denotes the charge-dispersion amplitude of the $k$-th energy level $\ket{k}$, $n_g(t)$ is the gate-controlled offset charge, $P\in\{+1,-1\}$ is the charge-parity, $\Omega_{\mathrm{WM}}(t)$ is driving strength of the Rabi oscillation.
Here, we employ the derivative removal by adiabatic gate to suppress population leakage to higher energy levels.





\begin{figure}
    \centering    \includegraphics[width=1\linewidth]{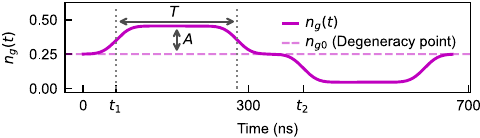} 
    \caption{The schematic waveform of a net-zero-based gate pulse.}
    \label{fig:ng(t)}
\end{figure}

For the gate drive term, we employ smoothed square pulses, which can both mitigate waveform distortion and enable faster, controllable accumulation of the parity-dependent phase. 
The pulse shape is defined as,
\[
f_{\mathrm{sq}}(t; t_0, T, \sigma)
= \tfrac{1}{2}\Big[
\operatorname{erf}\!\Big(\tfrac{t-t_0}{\sqrt{2}\sigma}\Big)
-
\operatorname{erf}\!\Big(\tfrac{t-t_0-T}{\sqrt{2}\sigma}\Big)
\Big],
\]
where $\operatorname{erf}(x)$ is the error function, providing smooth rising/falling edges, $t_0$ is the segment start time, $T$ is the flat-top duration, and $\sigma$ controls the edge smoothness. The net-zero-based gate pulse is constructed as follows,
\[
n_g(t)
=
n_{g0}
+
A\, f_{\mathrm{sq}}(t; t_1, T, \sigma)
-
A\, f_{\mathrm{sq}}(t; t_2, T, \sigma),
\]
where $n_{g0}$ represents the degeneracy point, $A$ is the gate pulse amplitude, and $t_1$ and $t_2$ are the start times of the first and second smoothed square pulse, respectively. Fig.~\ref{fig:ng(t)} shows a schematic waveform of the net-zero-based gate pulse.


\begin{table}[h]
\centering
\caption{Experimental parameters}
\label{tab:simulation_parameters}
\begin{tabular}{p{0.68\columnwidth}cc}
\hline
Parameter & Experiment \\
\hline
Qubit frequency $f_{01}$ & $3.51589~\text{GHz}$ \\
Dispersion energy difference $\epsilon_{10} /2\pi$& $1.192~\text{MHz}$ \\
Relaxation time $T_1$ & $80~\mu\text{s}$  \\
Dephasing time $T_{2}^\mathrm{Ramsey}$ ($n_{g}=0.25$)& $3.7\mu s$  \\

Dephasing time $T_{2}^\mathrm{Ramsey}$ ($n_{g}=0.5$)&$8.7\mu s$  \\ 
Dephasing time $T_{2}^\mathrm{echo}$ ($n_{g}=0.25$)& $40\mu s$ \\
Dephasing time $T_{2}^\mathrm{echo}$ ($n_{g}=0.5$)& $47\mu s$  \\
Anharmonicity $\eta/h$ & $0.33~\text{GHz}$ \\
Gate pulse amplitude $A$& $0.67~\text{V}$ \\
Duration of smoothed square pulse $T$&$217~\text{ns}$  \\
 Smoothness of square pulse $\sigma$& $5~\text{ns}$\\
 Total duration of smoothed square pulse&$237~\text{ns}$  \\

\hline
\end{tabular}
\end{table}

\begin{figure}[t]
    \centering   \includegraphics[width=1\linewidth]{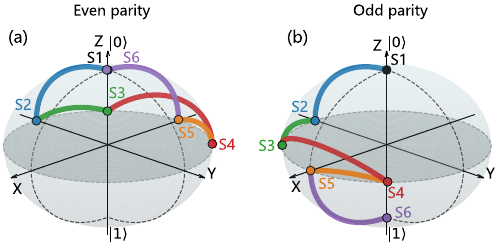}  
    \caption{Bloch-sphere trajectories of the charge-parity mapping process obtained from numerical simulations. (a) Even-parity. (b) Odd-parity. S1–S6 (circles) denote the quantum states after each pulse in the EchoCPM protocol, where S2-S5 lie on the equatorial plane.}
    \label{fig:Bloch-sphere}
\end{figure}

Table~\ref{tab:simulation_parameters} provides a summary of the experimental parameters. Using these well-calibrated parameters, we perform numerical simulations of the charge–parity mapping sequence. The evolution of the qubit state is visualized in the Bloch sphere in Fig.~\ref{fig:Bloch-sphere}, where the labeled states (S1–S6) correspond one-to-one with each control pulse depicted in Fig. 1(d) of the main text. 

Taking the even parity depicted in Fig.~\ref{fig:Bloch-sphere}(a) as an example, the evolution of the qubit state proceeds as follows. Starting from the ground state $\ket{0}$ (S1), an X/2 pulse is applied to the qubit, which manipulates the state to S2. Subsequently, the first smoothed square pulse induces a frequency shift of the qubit, leading to the qubit evolving to state S3. An X pulse is then employed to flip the qubit into state S4. Next, the second opposite smoothed square pulse evolves the qubit to state S5. Finally, a Y/2 pulse rotates the qubit to the final state S6. As shown in Figs.~\ref{fig:Bloch-sphere}(a) and (b), it can be seen that in the entire sequence, the even-parity and odd-parity branches arrive at opposite qubit states, respectively. To further validate the charge–parity mapping mechanism, we perform numerical simulations of the process using a Lindblad master-equation model. Here, we only consider the error rate due to relaxation and dephasing, with $T_1$ and $T_{2}^\mathrm{echo}$ parameters at $n_g = 0.5$ shown in the Table~\ref{tab:simulation_parameters}. Under these settings, this simulation yields an average charge-parity mapping fidelity of 99.36\%.




\section{ Model of Quasiparticle Tunneling}

Quasiparticle tunneling switches the charge parity of superconducting qubits, which can be modeled by random telegraph signals (RTS). We calculate the power spectrum density (PSD) of the RTS by considering infidelities of the charge-parity mapping and the qubit state readout. 

We denote the measured RTS at time t by P(t), where P(t) = +1 (-1) corresponds to the odd (even) charge parity of the qubit. The time interval between adjacent parity measurements is denoted by $\Delta$t. During the measurement, we assume a constant quasiparticle tunneling rate $\Gamma$. The autocorrelation function of the RTS is defined as $C(\tau)\equiv\langle P(t)P(t+\tau) \rangle$, where $\langle \cdot \rangle$ denotes the ensemble average. The PSD of the RTS, denoted by S(f), is the Fourier transform of $C(\tau)$,
\begin{equation}
S(f) \equiv \int_{-\infty}^{\infty} d\tau C(\tau) exp(-2\pi i f\tau).
\end{equation}
We first introduce two matrices to characterize the qubit readout and charge-parity mapping as follows,
\begin{equation}
\label{eq:Fshot}
F_{\mathrm{r}}=
\begin{bmatrix}
F_e   & 1-F_g\\
1-F_e & F_g
\end{bmatrix},\quad
F_{\mathrm{map}}=
\begin{bmatrix}
(F_m+1)/2 & (1-F_m)/2\\
(1-F_m)/2 & (F_m+1)/2
\end{bmatrix},
\end{equation}
where $F_g$ and $F_e$ are fidelities of the ground and excited states, $F_m$ is the fidelity of the charge-parity mapping.
Note that $F_m$ describes the probability of faithfully mapping the charge-parity states to the qubit states, while $1-F_m$ quantifies the probability of randomized mapping. This also corresponds to the fidelity measured using Randomized Benchmarking. Combining these two processes, the total fidelity matrix can be written as $F_{\text{tot}} = F_{\text{r}} F_{\text{map}}$.
Next, we introduce the vectors $\ket{e}$ and $\ket{o}$ to represent the even and odd parity states, and define the correlation matrix $M$ as follows,
\begin{equation}
M_{\rm{corr}} \equiv
\begin{pmatrix}
1 & -1\\
-1 & 1
\end{pmatrix},\qquad
\ket{e}\equiv\begin{pmatrix}1\\0\end{pmatrix},\qquad
\ket{o}\equiv\begin{pmatrix}0\\1\end{pmatrix}.
\label{eq:M}
\end{equation}
Then the expectation value of $P(t)P(t+\tau)$ can be calculated as, 
\begin{equation}
E_{\alpha\beta}=\mathbb{E}\!\left[P(t)P(t+\tau)\mid \alpha,\beta\right]
= \bra{\alpha}F_{\mathrm{tot}}^{\mathsf T} M_{\rm{corr}} F_{\mathrm{tot}}\ket{\beta},
\label{eq:Eab-def}
\end{equation}
where $\alpha,\beta\in\{e,o\}$ and $\ket{\alpha}$ represents $P(t)$ and $\ket{\beta}$ represents $P(t+\tau)$. The number of quasiparticle tunneling events $k$ follows a Poisson distribution with parameter $\lambda = \Gamma|\tau|$. When $k$ is even, the initial and final charge parity are the same, and when $k$ is odd, the initial and final charge parity are opposite.  Therefore, the autocorrelation function $C(\tau)=\langle P(t)P(t+\tau)\rangle$, when $\tau\neq0$, can be written as
\begin{equation}
\begin{aligned}
C(\tau\neq0)
&= \frac{E_{ee}+E_{oo}}{2}\cdot \sum_{k\,\text{even}} \frac{\lambda^k e^{-\lambda}}{k!}
+ \frac{E_{eo}+E_{oe}}{2}\cdot \sum_{k\,\text{odd}} \frac{\lambda^k e^{-\lambda}}{k!}\\
&= \frac{E_{ee}+E_{oo}}{2}\cdot \tfrac{1}{2}\bigl(1+e^{-2\lambda}\bigr)
+ \frac{E_{eo}+E_{oe}}{2}\cdot \tfrac{1}{2}\bigl(1-e^{-2\lambda}\bigr)\\
&=(F_e - F_g)^2 \;+\; F_{\mathrm{eff}}^{\,2}\, e^{-2\Gamma|\tau|}.
\end{aligned}
\end{equation}
where $F_{\mathrm{eff}} \;\equiv\; (F_e+F_g-1)\,F_m$. When $\tau = 0$, according to the definition, the correlation coefficient is 1. Therefore, 
\begin{equation}
\begin{aligned}
C(\tau)
&= (F_e - F_g)^2 \;+\; F_{\mathrm{eff}}^{\,2}\, e^{-2\Gamma|\tau|}\\
&+ \Bigl[1-F_{\mathrm{eff}}^{\,2}-(F_e-F_g)^{2}\Bigr]\delta (\tau=0),
\end{aligned}
\end{equation}
and the final PSD of the RTS takes the following expression,
\begin{equation}
\label{eq:Sf-Feff-final}
\begin{aligned}
S(f)&=\frac{F_{\mathrm{eff}}^{\,2}\,4\Gamma}{(2\Gamma)^{2}+(2\pi f)^{2}}+ \Bigl[1-F_{\mathrm{eff}}^{\,2}-(F_e-F_g)^{2}\Bigr]\\
&
\;+\;(F_e-F_g)^{2}\,\delta (f=0).
\end{aligned}
\end{equation}
In the discrete regime, where we define $\tau = n\Delta t$, $f_n = n/L$, $L$ is the total length of the RTS and obtain,
\begin{equation}
\label{eq:Sf-Feff-final}
S(f_n)=
\frac{F_{\mathrm{eff}}^{\,2}\,4\Gamma}{(2\Gamma)^{2}+(2\pi f_n)^{2}}
\;+\;\Bigl[1-F_{\mathrm{eff}}^{\,2}-(F_e-F_g)^{2}\Bigr]\Delta t.
\end{equation}

\section{ Ramsey-Based and echo-based charge-Parity detection}

\begin{figure}[!t]
     \centering  \includegraphics[width=1\linewidth]{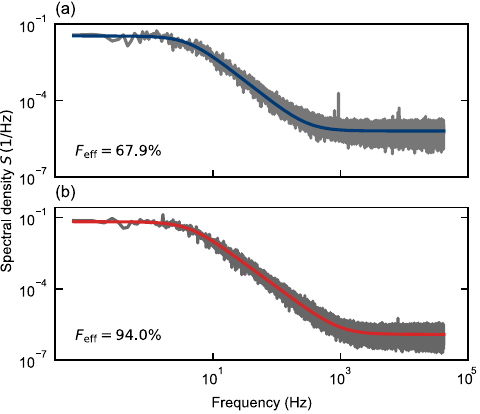}  
\caption{Comparison of PSD between (a) Ramsey-based and (b) Echo-based charge-parity detection. Both traces are fitted with the Eq.~\ref{eq:Sf-Feff-final}. }
    \label{fig:Ramsey_vs_echo}
\end{figure}

Ramsey and spin-echo protocols have long served as the primary methods for estimating qubit dephasing time. In the experiment, the dephasing time measured by the spin-echo protocol is typically longer than that of the Ramsey protocol. This is because the spin-echo protocol inserts a $\pi$ pulse between two $\pi/2$ pulses, thereby reversing and canceling out the phase error caused by low-frequency noise. Usually, the Ramsey protocol can be employed to detect phase accumulation. In the main text, we use the spin-echo protocol, combined with the net-zero-based gate pulse, to enable reliable detection of phase accumulation. To assess the two protocols, we perform experimental comparisons, with results detailed in Fig.~\ref{fig:Ramsey_vs_echo}. In our experiment, both measurements are conducted under identical environmental conditions, which ensures that any observed performance differences are attributed to the protocol itself. We analyze the PSD of the background QPs tunnelings and fit it using Eq.~\ref{eq:Sf-Feff-final}. We achieve effective detection fidelities of 67.9\% and 94.0\% for Ramsey-based and echo-based charge-parity detections, respectively. We attribute the unexpectedly poor performance of the standard Ramsey-based method to the shorter $T_2^\mathrm{Ramsey}$ time and electronic distortions on the gate-control lines. This highlights the advantage of using the EchoCPM protocol for charge-parity detection.

\bigskip
\noindent\textbf{AUTHOR CONTRIBUTIONS STATEMENT}

J.W. and X.L. conceived the experiment. X.L., T.S., Y.S., Y.-L.L., G.-M.X., W.-J.S., and M.-L.L. designed and fabricated the quantum device. Y.-Y.J., J.W., Y.-M.G., and Y.L. performed the measurement. J.W., Y.-Y.J., Y.L., Y.Z., and X.L. analyzed the experimental data. Y.-Y.J., J.W., and X.L. wrote the manuscript. H.-F.Y. and Y.-R.J. supervised the experiment. All authors discussed the results and the manuscript.

\bigskip
\noindent\textbf{COMPETING INTERESTS}

The authors declare no competing interests.

\end{document}